\documentclass[prd,twocolumn, amsmath,amssymb,floatfix,nofootinbib]{revtex4}
\usepackage{mathrsfs,bm}
\usepackage{longtable,lscape}
\usepackage{txfonts}
\usepackage{indentfirst}
\usepackage{graphicx,color,dcolumn,booktabs}
\usepackage{multirow}
\usepackage{indentfirst}
\usepackage{multirow}
\usepackage{epsfig}
\usepackage{graphicx,color,dcolumn}
\usepackage{epstopdf}
\usepackage{amsmath}
\usepackage{diagbox}
\usepackage{changes}
\usepackage{threeparttable}
\usepackage{booktabs}
\usepackage{color}
\usepackage{amssymb}
\usepackage{cancel}
\definecolor{cover}{rgb}{0.77,0.87,0.88}
\definecolor{blueone}{rgb}{0.1,0.1,.7}
\definecolor{citec}{rgb}{0.14,0.47,0.09}
\definecolor{two}{rgb}{0.0,0.5,0.}
\definecolor{three}{rgb}{.5,.1,0.15}
\usepackage[bookmarks=true,bookmarksopen=false,plainpages=false,breaklinks=true,
   bookmarksnumbered=true,hypertexnames=false,
   filecolor=blue,urlcolor=three,menucolor=three,
   linkcolor=three,citecolor=blueone, colorlinks,
   anchorcolor=blue,runcolor=pink,frenchlinks=red
   pdfstartview=FitH,pdftitle=title,%
   pdfauthor=author]{hyperref}

\usepackage{relsize}
\usepackage{xspace}
\def\babar{\mbox{\slshape B\kern-0.1em{\smaller A}\kern-0.1em
    B\kern-0.1em{\smaller A\kern-0.2em R}}}

\newcolumntype{C}{>{$}c<{$}}
\AtBeginDocument{
\heavyrulewidth=.08em
\lightrulewidth=.05em
\cmidrulewidth=.03em
\belowrulesep=.65ex
\belowbottomsep=0pt
\aboverulesep=.4ex
\abovetopsep=0pt
\cmidrulesep=\doublerulesep
\cmidrulekern=.5em
\defaultaddspace=.5em
}

\allowdisplaybreaks[4]

\begin{document}
\title{Pole trajectories from $S$- and $P$-wave $D\bar{D}^*$ interactions}
\author{Xiao-Xiao Chen, Zuo-Ming Ding, Jun He\footnote{Corresponding author: junhe@njnu.edu.cn}}

\affiliation{School of Physics and Technology, Nanjing Normal University, Nanjing 210097, China}

\date{\today}
\begin{abstract}

In this work, we investigate the $S$- and $P$-wave interactions of the
$D\bar{D}^*$ system within the framework of the quasipotential Bethe-Salpeter
equation, with the aim of exploring possible molecular states and their
corresponding pole trajectories. The interaction potentials are constructed
using the one-boson-exchange  model, incorporating the exchanges of $\pi$,
$\eta$, $\rho$, $\omega$, $\sigma$, and $J/\psi$ mesons, based on heavy quark
effective Lagrangians. The poles of the scattering amplitude are analyzed and their evolution on two Riemann sheets is systematically traced as the cutoff parameter increases up to 5 GeV.  We identify four molecular states arising
from the $S$-wave $D\bar{D}^*$ interaction. Among them, the bound state with
quantum numbers $I^G(J^{PC}) = 0^+(1^{++})$ corresponds well to the
experimentally observed $X(3872)$, while its isovector partner with $I^G(J^{PC})
= 1^-(1^{++})$ is found to exist only as a virtual state. Additionally, a
$0^-(1^{+-})$ state appears as a bound state. The isovector $1^{+-}$ state,
which may be associated with the $Z_c(3900)$, is observed to evolve from a bound
state to a virtual state as the interaction strength decreases. For the $P$-wave
$D\bar{D}^*$ interaction, the  structure $G(3900)$ recently observed at BESIII is likely
connected to a $0^-(1^{--})$ state. A $0^+(0^{-+})$ state is also predicted in
this channel. Both can appear as either resonance or bound/virtual state
depending on the interaction strength.

\end{abstract}

\maketitle

\section{Introduction}

Since 2003, an increasing number of XYZ particles have been observed in
experiments~\cite{Zyla:2020zbs,Olsen:2017bmm,Yuan:2018inv}. The discovery of
these exotic states has challenged the traditional quark model's classification
and interpretation of hadrons. Consequently, various theoretical frameworks have
been proposed to explore the origin and internal structure of the XYZ particles,
including molecular states, and compact multiquark
configurations, hadro-quarkonium~\cite{Voloshin:2007dx,Alberti:2016dru,Chen:2016qju,Wu:2021udi,Guo:2017jvc,Ali:2017jda,Dong:2017gaw,Guo:2019twa,Cleven:2015era,Wang:2020dgr}.
Among these, the molecular state interpretation has gained particular attention,
especially given that many observed XYZ states lie close to hadronic
thresholds~\cite{Zyla:2020zbs}.

In the original molecular picture, a molecular state is regarded as a loosely
bound system of two hadrons, analogous to the deuteron~\cite{Tornqvist:1993ng}. The $X(3872)$ was the
first XYZ particle to be experimentally discovered, and its proximity to the
$D\bar{D}^*$ threshold has led to widespread interpretation as a weakly bound
$D\bar{D}^*$ molecular
state~\cite{Choi:2003ue,Lee:2009hy,Tornqvist:2004qy,Close:2003sg,Barnes:2003vb,Kalashnikova:2005ui}.
In Ref.~\cite{Li:2012cs},  the
one-boson-exchange model was employed to investigate the possibility of
$X(3872)$ being a $D\bar{D}^*$ molecular state with quantum numbers $J^{PC} =
1^{++}$. The observation of hidden-charm pentaquarks at LHCb further supports
this picture, as they form a rich spectrum of $S$-wave molecular states with
masses  slightly  below the corresponding thresholds.  However, the discovery of states,
such as $Z_c(3900)$, $Z_b(10610)$, and $Z_b(10650)$, poses a challenge to this
simple molecular
scenario~\cite{Lin:2024qcq,Liu:2013dau,Xiao:2013iha,Wang:2013cya,Wang:2013daa,Zhang:2013aoa,Aceti:2014uea,He:2014nya}.
Although their masses lie close to the $D\bar{D}^*$ and $B^{(*)}\bar{B}^*$
thresholds, respectively, they are slightly above the thresholds, which makes it
difficult to interpret them as loosely bound molecular states in the traditional
sense.

According to standard scattering theory, a bound state generated from a
single-channel interaction appears as a pole on the real axis below the
threshold in the complex energy plane. In addition to bound states, virtual
states located on the real axis below threshold but on the second Riemann sheet
and resonances positioned above the threshold in the complex plane can also
arise~\cite{Taylor:1972pty}. All three types of states, bound, virtual, and
resonance, stem from attractive interactions; however, forming a bound state
generally requires a stronger attraction.  A representative example is the
$\Lambda(1405)$, which is widely interpreted as a dynamically generated state
with a two-pole structure. Notably, the higher-energy $\bar{K}N$ channel exhibits
stronger attraction, sufficient to support a bound state, whereas the
lower-energy $\pi\Sigma$ channel has relatively weaker attraction, though still
strong enough to generate a resonance~\cite{Hyodo:2011ur,He:2015cca}.  In
Refs.~\cite{Du:2022jjv,He:2017lhy}, detailed fits to experimental data suggest
that the $Z_c(3900)$ may originate from a virtual state. Although a virtual
state lies below the threshold, its location on the second Riemann sheet allows
it to produce observable enhancements above the threshold, rather than below.
In Ref~\cite{Dong:2021juy}, a survey of heavy-antiheavy hadronic molecules was
also performed, considering both bound and virtual states. Through a
simultaneous analysis of BESIII data on the processes $e^+e^- \to \gamma
(D^0\bar{D}^{*0}\pi^0 / J/\psi \pi^+\pi^-)$ and LHCb data on $B^+ \to K^+
(J/\psi \pi^+ \pi^-)$, an isovector partner of the $X(3872)$ has also been
identified as a virtual state in Ref.~\cite{Ji:2025hjw}. These findings further
support the idea that many near-threshold structures, rather than being bound
states, may instead be virtual states even resonance arising from strong but
insufficiently attractive interactions.

In contrast to the well-established $S$-wave molecular states, the existence of
$P$-wave molecular states remains a topic of ongoing debate. A possible new
structure, referred to as $G(3900)$, was recently observed by the BESIII
Collaboration through high-precision measurements of the Born cross section in
the process $e^+e^- \to D\bar{D}$~\cite{BESIII:2024ths}, which was also observed at the B factories~\cite{BaBar:2006qlj,Belle:2007qxm}. This structure lies near
the $D\bar{D}^*$ threshold but possesses spin parity $J^P = 1^-$,
making it a strong candidate for a $P$-wave molecular state. Recent theoretical
studies and model calculations suggest that $G(3900)$ can be interpreted as the
first $P$-wave $D\bar{D}^*$ resonance~\cite{Lin:2024qcq,Huang:2025rvj}.  In
addition, the possible existence of other $D\bar{D}^*$ molecular states with
spin parities $0^-$ and $1^-$ has also been investigated in the
literature~\cite{Lin:2024qcq}.

In this work, we carry out a systematic investigation of possible molecular states arising from both $S$- and $P$-wave interactions of the $D\bar{D}^*$ system, including those potentially related to the newly observed $G(3900)$ and the isovector partner of the $X(3872)$. Our previous studies have primarily focused on analyzing $D\bar{D}^*$ interactions in the $S$-wave channel to identify molecular candidates and explore their correspondence with experimentally observed states~\cite{He:2014nya,Ding:2020dio,Sun:2012zzd,He:2015mja,He:2017lhy}. However, the trajectories of scattering amplitude poles generated by such interactions have not been thoroughly examined and remain scarcely discussed in the existing literature.

The study of pole trajectories, i.e., how poles in the complex energy plane evolve with variations in interaction parameters, can provide deeper insight into the internal structures of near-threshold states. As demonstrated in Ref.~\cite{Hanhart:2014ssa}, analyzing the movement of poles as a function of interaction strength is particularly informative for distinguishing $S$-wave and $P$-wave dynamics and for identifying virtual, bound, or resonant states.
Motivated by this, we systematically explore the pole trajectories associated
with $D\bar{D}^*$ interactions using the quasipotential Bethe-Salpeter equation
(qBSE) approach in combination with a one-boson-exchange (OBE) model. To reduce
complexity, the analysis is restricted to a single-channel framework. By varying
the cutoff parameter, we search for poles corresponding to different
isospins, $G$-parities, and spin parities (corresponding to $S$- and
$P$-waves) in the $D\bar{D}^*$ system.

This article is organized as follows. Following the introduction,
Section~\ref{Sec: Formalism} presents the theoretical framework, including the
flavor wave functions, effective Lagrangians, construction of the interaction
potentials, and a brief overview of the qBSE approach. The numerical results,
with particular focus on the pole trajectories arising from the $S$- and
$P$-wave $D\bar{D}^*$ interactions, are provided in Section~\ref{Sec: Results}.
Finally, Section~\ref{Sec: Summary} summarizes our findings and discusses their
implications for the understanding of near-threshold exotic states.

\section{Theoretical frame} \label{Sec: Formalism}

First, we construct the flavor wave function of $D\bar{D}^{*}$ state under SU (3) symmetry as~\cite{He:2014nya,Sun:2011uh},
\begin{align}
|X_{D\bar{D}^*}^0\rangle_{I=0}&=\frac{1}{2}\Big[\big(|D^{*+}D^-\rangle+|D^{*0}\bar{D}^0\rangle\big)+c\big(|D^+D^{*-}\rangle+|D^0\bar{D}^{*0}\rangle\big)\Big], \nonumber\\
|X_{D\bar{D}^*}^0\rangle_{I=1}&=\frac{1}{2}\Big[\big(|D^{*+}D^-\rangle-|D^{*0}\bar{D}^0\rangle\big)+c\big(|D^+D^{*-}\rangle-|D^0\bar{D}^{*0}\rangle\big)\Big],\nonumber\\
|X_{D\bar{D}^*}^+\rangle_{I=1}&=\frac{1}{\sqrt{2}}\big(|D^{*+}\bar{D}^0\rangle+c|D^+\bar{D}^{*0}\rangle\big),\nonumber\\
|X_{D\bar{D}^*}^-\rangle_{I=1}&=\frac{1}{\sqrt{2}}\big(|D^{*-}\bar{D}^0\rangle+c|D^-\bar{D}^{*0}\rangle\big),
 \label{Eq: wf1}
\end{align}
where $c=\pm$ corresponds to $C$ parity $C=\mp$ respectively. $C$ parity cannot be defined in the isovector state, and $G$ parity is used instead.

We use the one-boson-exchange model to construct the interaction potential, which includes the $\pi$, $\eta$, $\rho$, $\omega$ and $\sigma$ exchanges. Considering the chiral symmetry, we first give the Lagrangian for the coupling vertices of heavy meson-heavy meson and exchange light meson as follows~\cite{Cheng:1992xi,Yan:1992gz,Wise:1992hn,Burdman:1992gh,Casalbuoni:1996pg},
\begin{align}
\mathcal{L}_{\mathcal{P}^*\mathcal{P}^*\mathbb{P}} &=
-i\frac{2g}{f_\pi}\varepsilon_{\alpha\mu\nu\lambda}
v^\alpha\mathcal{P}^{*\mu}_{b}{\mathcal{P}}^{*\lambda\dag}_{a}
\partial^\nu{}\mathbb{P}_{ba}\nonumber\\
&+i \frac{2g}{f_\pi}\varepsilon_{\alpha\mu\nu\lambda}
v^\alpha\widetilde{\mathcal{P}}^{*\mu\dag}_{a}\widetilde{\mathcal{P}}^{*\lambda}_{b}
\partial^\nu{}\mathbb{P}_{ab},\nonumber\\
\mathcal{L}_{\mathcal{P}^*\mathcal{P}\mathbb{P}} &=-
\frac{2g}{f_\pi}(\mathcal{P}^{}_b\mathcal{P}^{*\dag}_{a\lambda}+
\mathcal{P}^{*}_{b\lambda}\mathcal{P}^{\dag}_{a})\partial^\lambda{}
\mathbb{P}_{ba}\nonumber\\
&+\frac{2g}{f_\pi}(\widetilde{\mathcal{P}}^{*\dag}_{a\lambda}\widetilde{\mathcal{P}}_b+
\widetilde{\mathcal{P}}^{\dag}_{a}\widetilde{\mathcal{P}}^{*}_{b\lambda})\partial^\lambda{}\mathbb{P}_{ab}.
\nonumber\\
  \mathcal{L}_{\mathcal{PP}\mathbb{V}}
  &= -\sqrt{2}\beta{}g_V\mathcal{P}^{}_b\mathcal{P}_a^{\dag}
  v\cdot\mathbb{V}_{ba}
 +\sqrt{2}\beta{}g_V\widetilde{\mathcal{P}}^{\dag}_a
  \widetilde{\mathcal{P}}^{}_b
  v\cdot\mathbb{V}_{ab},\nonumber\\
  \mathcal{L}_{\mathcal{P}^*\mathcal{P}\mathbb{V}}
  &=- 2\sqrt{2}\lambda{}g_V v^\lambda\varepsilon_{\lambda\mu\alpha\beta}
  (\mathcal{P}^{}_b\mathcal{P}^{*\mu\dag}_a +
  \mathcal{P}_b^{*\mu}\mathcal{P}^{\dag}_a)
  (\partial^\alpha{}\mathbb{V}^\beta)_{ba}\nonumber\\
&-  2\sqrt{2}\lambda{}g_V
v^\lambda\varepsilon_{\lambda\mu\alpha\beta}
(\widetilde{\mathcal{P}}^{*\mu\dag}_a\widetilde{\mathcal{P}}^{}_b
+
\widetilde{\mathcal{P}}^{\dag}_a\widetilde{\mathcal{P}}_b^{*\mu})
  (\partial^\alpha{}\mathbb{V}^\beta)_{ab},\nonumber\\
  \mathcal{L}_{\mathcal{P}^*\mathcal{P}^*\mathbb{V}}
  &= \sqrt{2}\beta{}g_V \mathcal{P}_b^{*}\cdot\mathcal{P}^{*\dag}_a
  v\cdot\mathbb{V}_{ba}\nonumber\\
  &-i2\sqrt{2}\lambda{}g_V\mathcal{P}^{*\mu}_b\mathcal{P}^{*\nu\dag}_a
  (\partial_\mu{}
  \mathbb{V}_\nu - \partial_\nu{}\mathbb{V}_\mu)_{ba}\nonumber\\
  &-\sqrt{2}\beta g_V
  \widetilde{\mathcal{P}}^{*\dag}_a\widetilde{\mathcal{P}}_b^{*}
  v\cdot\mathbb{V}_{ab}\nonumber\\
  &-i2\sqrt{2}\lambda{}g_V\widetilde{\mathcal{P}}^{*\mu\dag}_a\widetilde{\mathcal{P}}^{*\nu}_b(\partial_\mu{}
  \mathbb{V}_\nu - \partial_\nu{}\mathbb{V}_\mu)_{ab}.
\nonumber\\
  \mathcal{L}_{\mathcal{PP}\sigma}
  &= -2g_s\mathcal{P}^{}_b\mathcal{P}^{\dag}_b\sigma
 -2g_s\widetilde{\mathcal{P}}^{}_b\widetilde{\mathcal{P}}^{\dag}_b\sigma,\nonumber\\
  \mathcal{L}_{\mathcal{P}^*\mathcal{P}^*\sigma}
  &= 2g_s\mathcal{P}^{*}_b\cdot{}\mathcal{P}^{*\dag}_b\sigma
 +2g_s\widetilde{\mathcal{P}}^{*}_b\cdot{}\widetilde{\mathcal{P}}^{*\dag}_b\sigma,\label{Eq:L}
\end{align} 
where  the velocity $v$ should be replaced by $i\overleftrightarrow{\partial}/2\sqrt{m_im_f}$ with the $m_{i,f}$ being the mass of the initial or final heavy meson. 
${\mathcal{P}}^{(*)T} =(D^{(*)0},D^{(*)+},D_s^{(*)+})$ and
$\widetilde{\mathcal{P}}^{(*)T}=(\bar{D}^{(*)0},\bar{D}^{(*)-},\bar{D}_s^{(*)-})$, and 
 satisfy the normalization relations $\langle
0|{\mathcal{P}}|{Q}\bar{q}(0^-)\rangle
=\sqrt{M_\mathcal{P}}$ and $\langle
0|{{\mathcal{P}}}^*_\mu|{Q}\bar{q}(1^-)\rangle=
\epsilon_\mu\sqrt{M_{\mathcal{P}^*}}$. 
The pseudoscalar and the $\mathbb P$ and $\mathbb V$ are the pseudoscalar and vector matrices
\begin{equation}
    {\mathbb P}=\left(\begin{array}{ccc}
        \frac{\sqrt{3}\pi^0+\eta}{\sqrt{6}}&\pi^+&K^+\\
        \pi^-&\frac{-\sqrt{3}\pi^0+\eta}{\sqrt{6}}&K^0\\
        K^-&\bar{K}^0&-\frac{2\eta}{\sqrt{6}}
\end{array}\right),
\mathbb{V}=\left(\begin{array}{ccc}
\frac{\rho^0+\omega}{\sqrt{2}}&\rho^{+}&K^{*+}\\
\rho^{-}&\frac{-\rho^{0}+\omega}{\sqrt{2}}&K^{*0}\\
K^{*-}&\bar{K}^{*0}&\phi
\end{array}\right).\label{MPV}
\end{equation}

In this Lagrangians, coupling constants—$g_s$, $g$, $\beta$, and $\lambda$—are invovled, along with the pion decay constant $f_\pi = 132$ MeV and the parameter $g_V = m_\rho / f_\pi = 5.9$. Specifically, due to spontaneously broken chiral symmetry~\cite{Bardeen:2003kt}, $g_s$ is related to the coupling $\tilde{g}$ for the process $D(0^+) \rightarrow D(0^-) \pi$ via $g_s = \tilde{g} / (2\sqrt{6})$, with $\tilde{g} = 3.73$ taken from Ref.~\cite{Falk:1992cx}. The $D^* D^* \pi$ coupling constant is assumed to be equal to that of the $D^* D \pi$ vertex, $g = 0.59$, which is extracted from the decay width of $D^{*+}$~\cite{Isola:2003fh}. The value of $\beta = 0.9$ is fixed based on vector meson dominance~\cite{Isola:2003fh}, and $\lambda = 0.56$ GeV$^{-1}$ is obtained by comparing form factors from light-cone sum rules and lattice QCD~\cite{Isola:2003fh}. These values are widely accepted in the literature and have been adopted in many phenomenological studies~\cite{Liu:2008tn,Chen:2019asm}. For clarity, we have listed them in Table~\ref{coupling}.

This work also considers the couplings of heavy-light charmed mesons to $J/\psi$, and writes the correlation Lagrangians as~\cite{Casalbuoni:1996pg,Oh:2000qr},
\begin{eqnarray}
	{\cal L}_{D^*_{(s)}\bar{D}^*_{(s)}J/\psi}&=&-ig_{D^*_{(s)}D^*_{(s)}\psi}\big[\psi \cdot \bar{D}^*\overleftrightarrow{\partial}\cdot D^*\nonumber\\
&-&
\psi^\mu \bar D^* \cdot\overleftrightarrow{\partial}^\mu {D}^* +
\psi^\mu \bar{D}^*\cdot\overleftrightarrow{\partial} D^{*\mu} ) \big], \nonumber \\
{\cal L}_{D_{(s)}^*\bar{D}_{(s)}J/\psi}&=&
g_{D^*_{(s)}D_{(s)}\psi} \,  \, \epsilon_{\beta \mu \alpha \tau}
\partial^\beta \psi^\mu (\bar{D}
\overleftrightarrow{\partial}^\tau D^{* \alpha}+\bar{D}^{* \alpha}
\overleftrightarrow{\partial}^\tau D) \label{matrix3}, \nonumber \\
{\cal L}_{D_{(s)} \bar{D}_{(s)}J/\psi} &=&
ig_{D_{(s)}D_{(s)}\psi} \psi \cdot
\bar{D}\overleftrightarrow{\partial}D,
\end{eqnarray}
where the couplings
are related to a single parameter $g_2$ as
\begin{eqnarray}
\frac{g_{D^*D^*\psi} }{m_{D^*}}= \frac{g_{D_{(s)}D_{(s)}\psi}}{m_D}= g_{D^*_{(s)}D_{(s)}\psi}= 2 g_2 \sqrt{m_\psi },\label{Eq: para}
\end{eqnarray}
with $g_2={\sqrt{m_\psi}}/({2m_Df_\psi})$. The coupling constants used in the calculation are shown in Table~\ref{coupling}  below.

\renewcommand\tabcolsep{0.37cm}
\renewcommand{\arraystretch}{1.8}
\begin{table}[h!]
\caption{The coupling constants adopted in the
calculation, which are cited from the literature~\cite{Falk:1992cx,Isola:2003fh,Liu:2008tn,Chen:2019asm}. The $\lambda$ is in the units of GeV$^{-1}$. The $f_{\pi,\psi}$ are in the units of MeV. Others are in the units of $1$.
\label{coupling}}
\begin{tabular}{cccccccccccccccccc}\toprule[2pt]
$\beta$&$g$&$g_V$&$\lambda$ &$g_{s}$&$f_\pi$&$f_\psi$\\\hline
$0.9$&$0.59$&$5.9$&$0.56$ &$0.76$&$132$&$405$\\
\bottomrule[2pt]
\end{tabular}
\end{table}

With the above Lagrangians, the OBE interaction potentials can be constructed
using standard Feynman rules as
follows~\cite{He:2014nya,He:2015mja,He:2017lhy,He:2019ify}: \begin{equation}
{\cal V}_{\mathbb{P},\sigma} = I^{(d,c)}_i \Gamma_1 \Gamma_2
P_{\mathbb{P},\sigma} f(q^2), {\cal V}_{\mathbb{V}} = I^{(d,c)}_i \Gamma_{1\mu}
\Gamma_{2\nu} P^{\mu\nu}_{\mathbb{V}} f(q^2), \label{V} \end{equation} where
$\Gamma_1$ and $\Gamma_2$ denote the vertices at the upper and lower ends of the
exchanged boson, respectively. The explicit expressions for these vertices can
be found in our previous works~\cite{Ding:2020dio}.  To incorporate the
off-shell effects of the exchanged mesons and the extended structure of the
interaction vertices, a monopole form factor is introduced: $f(q^2) =
\Lambda_e^2 / (q^2 - \Lambda_e^2)$, as adopted in
Refs.~\cite{He:2015mja,Gross:1991pm}. Physically, the form factor introduced at
each vertex encodes the compositeness of heavy mesons when probed by exchange 
mesons, reflecting the finite size effects of the hadrons involved. Its precise
functional form and the associated cutoff parameter $\Lambda$ are not well
constrained by either experimental data or first-principles calculations. As is
commonly done in OBE  models, we treat $\Lambda$ as a free
parameter that can absorb certain model-dependent uncertainties. All couplings
in the OBE potential are taken from effective Lagrangians with
phenomenologically fixed constants and are independent of
$\Lambda$. In our framework, the short-range part of the interaction is modeled
by including exchanges of heavier mesons such as $\rho$, $\omega$, and $J/\psi$,
rather than by introducing explicit local four-point contact terms, as typically
done in effective field theory. The omission of such contact
interactions—whose couplings are in general cutoff-dependent—implies that our
cutoff variation only partially probes the full parameter space of the model.
Nevertheless, it is expected that this approach provides a reasonable first approximation to the short-range dynamics within the limitations of the OBE framework.

When discussing the interaction of double heavy flavor state $D\bar{D}^*$, the coefficients are collected as flavor factors. In Table~\ref{flavor factor},  flavor factors $I^d_i$ and $I^c_i$ of certain meson exchange $i$ of certain interaction are listed for direct and cross diagrams, respectively.
\renewcommand\tabcolsep{0.108cm}
\renewcommand{\arraystretch}{2}
\begin{table}[h!]
\begin{center}
\caption{The isospin factors $I_i^d$ and $I_i^c$ for direct and cross diagrams and different exchange mesons. 
\label{flavor factor}}
\begin{tabular}{c|cccccc|ccccccc}\bottomrule[2pt]
 & \multicolumn{6}{c|}{$I_i^d$}& \multicolumn{6}{c}{ $I_i^c$}\\\hline
&$\pi$&$\eta$  &$\rho$ &$\omega$&$\sigma$ &$J/\psi$ &$\pi$&$\eta$  &$\rho$ &$\omega$&$\sigma$ &$J/\psi$ \\\hline
$[{D}\bar{{D}}^{*}]^T$&$\cdots$&$\cdots$&$-\frac{1}{2}$ &$\frac{1}{2}$ &$1$ &$1$
&$-\frac{1}{2}c$&$\frac{1}{6}c$&$-\frac{1}{2}c$ &$\frac{1}{2}c$ & $\cdots$ &$c$\\
$[{D}\bar{{D}}^{*}]^S$&$\cdots$&$\cdots$&$\frac{3}{2}$&$\frac{1}{2}$&$1$ &$1$
&$\frac{3}{2}c$&$\frac{1}{6}c$&$\frac{3}{2}c$ &$\frac{1}{2}c$ & $\cdots$ &$c$\\
\toprule[2pt]
\end{tabular}
\end{center}
\end{table}

In the Bethe-Salpeter equation, the potential kernel is the central part that describes the interaction between two particles. The scattering amplitude can be obtained by inserting potential kernel into the Bethe-Salpeter equation. Through quasipotential approximation and partial wave decomposition, the four-dimensional integral equation of Minkowski space is reduced to a one-dimensional equation with spin parity $J^P$, as follows
~\cite{He:2014nya,He:2015mja,He:2017lhy,He:2015yva,He:2017aps},
\begin{align}
i{\cal M}^{J^P}_{\lambda'\lambda}({\rm p}',{\rm p})
&=i{\cal V}^{J^P}_{\lambda',\lambda}({\rm p}',{\rm
p})+\sum_{\lambda''}\int\frac{{\rm
p}''^2d{\rm p}''}{(2\pi)^3}\nonumber\\
&\cdot
i{\cal V}^{J^P}_{\lambda'\lambda''}({\rm p}',{\rm p}'')
G_0({\rm p}'')i{\cal M}^{J^P}_{\lambda''\lambda}({\rm p}'',{\rm
p}),\quad\quad \label{Eq: BS_PWA}
\end{align}
where the summation is limited to nonnegative helicities $\lambda''$.The propagator $G_0({\rm p}'')$ is reduced from its original four-dimensional form using the quasipotential approximation and takes the form
\begin{align}
G_0 &= \frac{\delta^+(p''^{2}_h - m_h^{2})}{p''^{2}_l - m_l^{2}} \nonumber\\
&= \frac{\delta^+(p''^{0}_h - E_h({\rm p}''))}{2E_h({\rm p}'')[(W - E_h({\rm p}''))^2 - E_l^2({\rm p}'')]}.
\end{align}
Following the spectator approximation adopted in this work, the heavier particle (denoted as $h$) in a given channel is placed on shell~\cite{Gross:1999pd}, satisfying $p''^0_h = E_h({\rm p}'') = \sqrt{m_h^2 + {\rm p}''^2}$. The energy of the lighter particle (denoted as $l$) is then given by $p''^0_l = W - E_h({\rm p}'')$, where $W$ is the total energy in the center-of-mass frame. Here and throughout the paper, the three-momentum in the center-of-mass frame is defined as ${\rm p} = |{\bm p}|$. The partial wave potential can be expressed by the potential in Eq.~(\ref{V}) as follows
\begin{align}
{\cal V}_{\lambda'\lambda}^{J^P}({\rm p}',{\rm p})
&=2\pi\int d\cos\theta
~[d^{J}_{\lambda\lambda'}(\theta)
{\cal V}_{\lambda'\lambda}({\bm p}',{\bm p})\nonumber\\
&+\eta d^{J}_{-\lambda\lambda'}(\theta)
{\cal V}_{\lambda'-\lambda}({\bm p}',{\bm p})],
\end{align}
where $\eta = PP_1P_2(-1)^{J-J_1-J_2}$. $P$ and $P_1$,$P_2$ represent the parity of the system and the parity of the two constituent heavy mesons, respectively. $J$ and $J_1$,$J_2$ represent the total spin of the system and the spins of the two constituent heavy mesons, respectively. The initial and final relative momenta are chosen as ${\bm p}=(0,0,{\rm p})$  and ${\bm p}'=({\rm p}'\sin\theta, 0, {\rm p}'\cos\theta)$. $d^J_{\lambda\lambda'}(\theta)$ is the Wigner d-matrix, which describes the rotation relationship between the initial and final relative momenta. 
An exponential regularization was also introduced as a form factor in the reduced propagator as $G_0({\rm p}'')\to G_0({\rm p}'')e^{-2(p''^2_l-m_l^2)^2/\Lambda_r^4}$~\cite{He:2015mja}.

\section{Pole trajectory of $D\bar{D}^*$ interaction}\label{Sec: Results}

Based on the above theoretical framework, the scattering amplitude for the $D\bar{D}^*$ interaction is reduced to a one-dimensional integral equation. Molecular states are identified as poles in the complex energy plane by solving the condition $|1 - V(z)G(z)| = 0$, where $z = E_R - i\Gamma/2$ denotes the complex energy~\cite{He:2015mja,Gross:1991pm}. For simplicity, we set the two form factor cutoff parameters, $\Lambda_e$ and $\Lambda_r$, equal and denote them collectively as a single cutoff parameter $\Lambda$.
Since the coupling constants are fixed, we treat $\Lambda$ as the only free parameter and investigate how the pole trajectories evolve with its variation. This effectively modulates the interaction strength, as a larger $\Lambda$ generally leads to a stronger attractive potential. Although the physical value of $\Lambda$ is unknown, scanning over a range allows us to assess the qualitative behavior and robustness of the predictions.
We vary $\Lambda$ up to 5~GeV on two Riemann sheets and focus on poles within 50~MeV of the $D\bar{D}^*$ threshold. All relevant isospin, spin, parity, and $G$-parity quantum numbers arising from $S$- and $P$-wave interactions are considered. Higher partial waves such as $D$ and $F$ waves may also contribute to the same spin-parity channels due to the structure of the partial wave decomposition. Only cases that produce bound states, virtual states, or resonances are included in the results.

\subsection{S-wave pole trajectory}

\begin{figure}[h!]
\includegraphics[bb=0 0 1000 520,clip,scale=0.335]{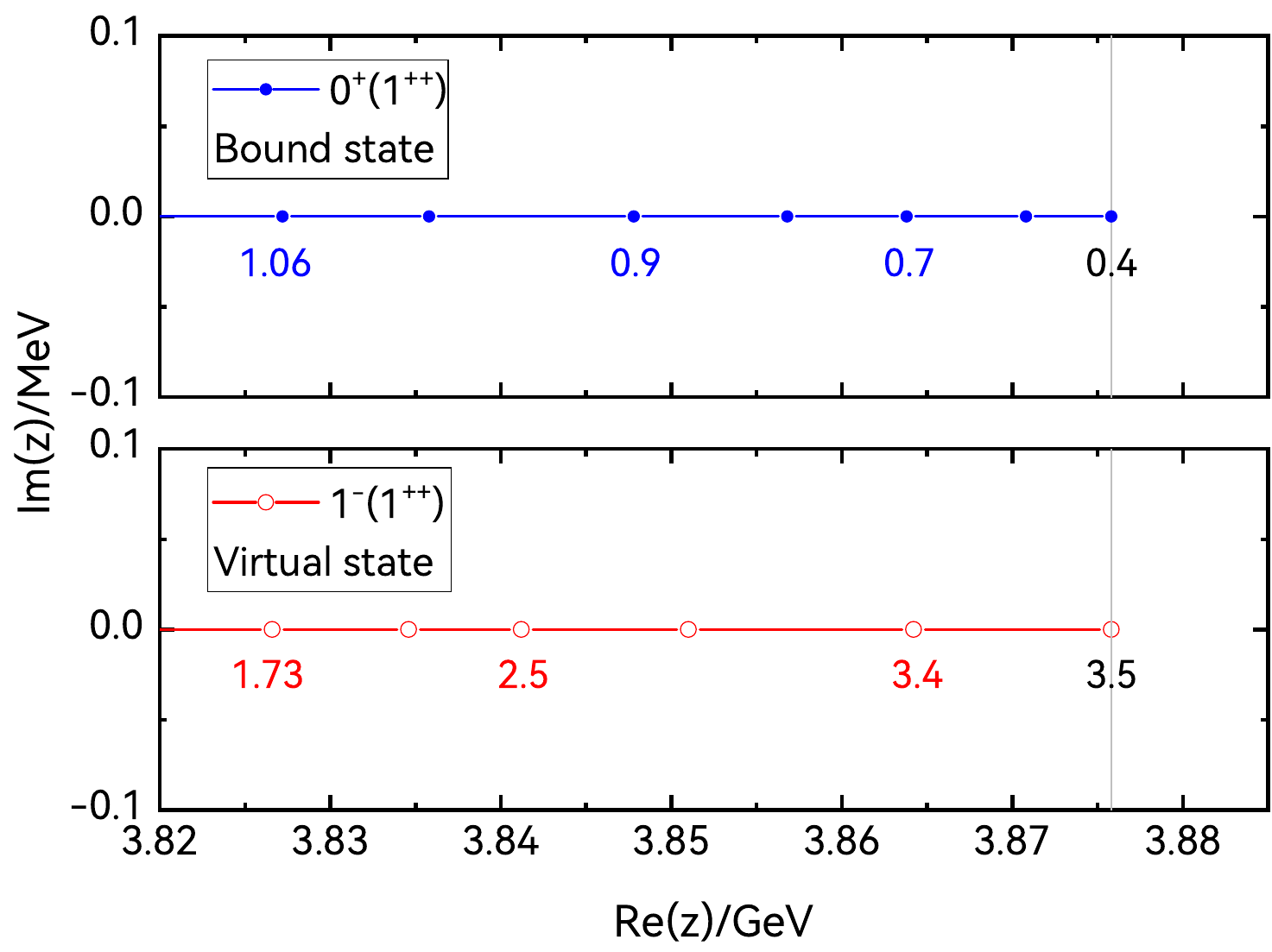}
\caption{Pole trajectories for the isoscalar (upper panel) and isovector (lower panel) states with quantum numbers $1^{++}$. For the $0^{+}(1^{++})$ channel, the corresponding cutoff values from left to right are 1.06, 1.0, 0.9, 0.8, 0.7, 0.6, and 0.4~GeV. For the $1^{-}(1^{++})$ channel, the cutoff values are 1.73, 2.0, 2.5, 3.2, 3.4, and 3.5~GeV. The gray solid line indicates the $D\bar{D}^*$ threshold.}
\label{1++}
\end{figure}

For the $S$-wave interaction, poles are found in the channels with quantum
numbers $I^G(J^{PC}) = 0^{+}(1^{++})$, $1^{-}(1^{++})$, $0^{-}(1^{+-})$, and
$1^{+}(1^{+-})$.  The upper panel of Fig.\ref{1++} illustrates the evolution of
the pole positions for the isoscalar state with $1^{++}$. When the cutoff
parameter $\Lambda$ is around 0.4~GeV, the pole is located at the threshold. As
the interaction becomes stronger with increasing $\Lambda$, the pole moves below
the threshold along the real axis on the first Riemann sheet, reaching
approximately 50~MeV below threshold at $\Lambda = 1.06$~GeV. With further
increase of the cutoff, the pole continues to move deeper, but we limit our
analysis to bound states within 50~MeV below the threshold. Although the cutoff
is already quite small and further reduction may lack direct physical meaning,
we nevertheless examine its behavior for completeness. Under such conditions, no
virtual state appears on the second Riemann sheet. Therefore, for the
$0^{+}(1^{++})$ channel, only bound states exist, with no corresponding virtual
state or resonance observed.  This isoscalar bound state with $I^G(J^{PC}) =
0^{+}(1^{++})$ can be naturally associated with the experimentally observed
$X(3872)$, in agreement with previous
studies~\cite{LHCb:2013kgk,Swanson:2003tb}.

As shown in the lower panel of Fig.~\ref{1++}, unlike the isoscalar case, only a
virtual state is found for the isovector channel with quantum numbers
$1^{-}(1^{++})$. When the cutoff $\Lambda$ is around 3.5~GeV, the pole appears
near the threshold. As $\Lambda$ decreases, the pole moves along the real axis
on the second Riemann sheet, reaching approximately 50~MeV below the threshold
when the cutoff is reduced to around 1.8~GeV, and continues to shift further as
the cutoff decreases. As the interaction weakens, the pole is unable to enter
the first Riemann sheet to form a bound state. Several recent
studies~\cite{Zhang:2024fxy,Dias:2024zfh,Sadl:2024dbd} have predicted the
existence of an isovector charmonium-like $D\bar{D}^*$ hadronic molecule with
$1^{-}(1^{++})$. In particular, the analysis in Ref.~\cite{Ji:2025hjw} suggests
that such a state should manifest as a virtual state. Our results support this
prediction, indicating the presence of a $1^{-}(1^{++})$ molecular state as a
virtual state.

\begin{figure}[h!]
\includegraphics[bb=0 0 1000 710,clip,scale=0.335]{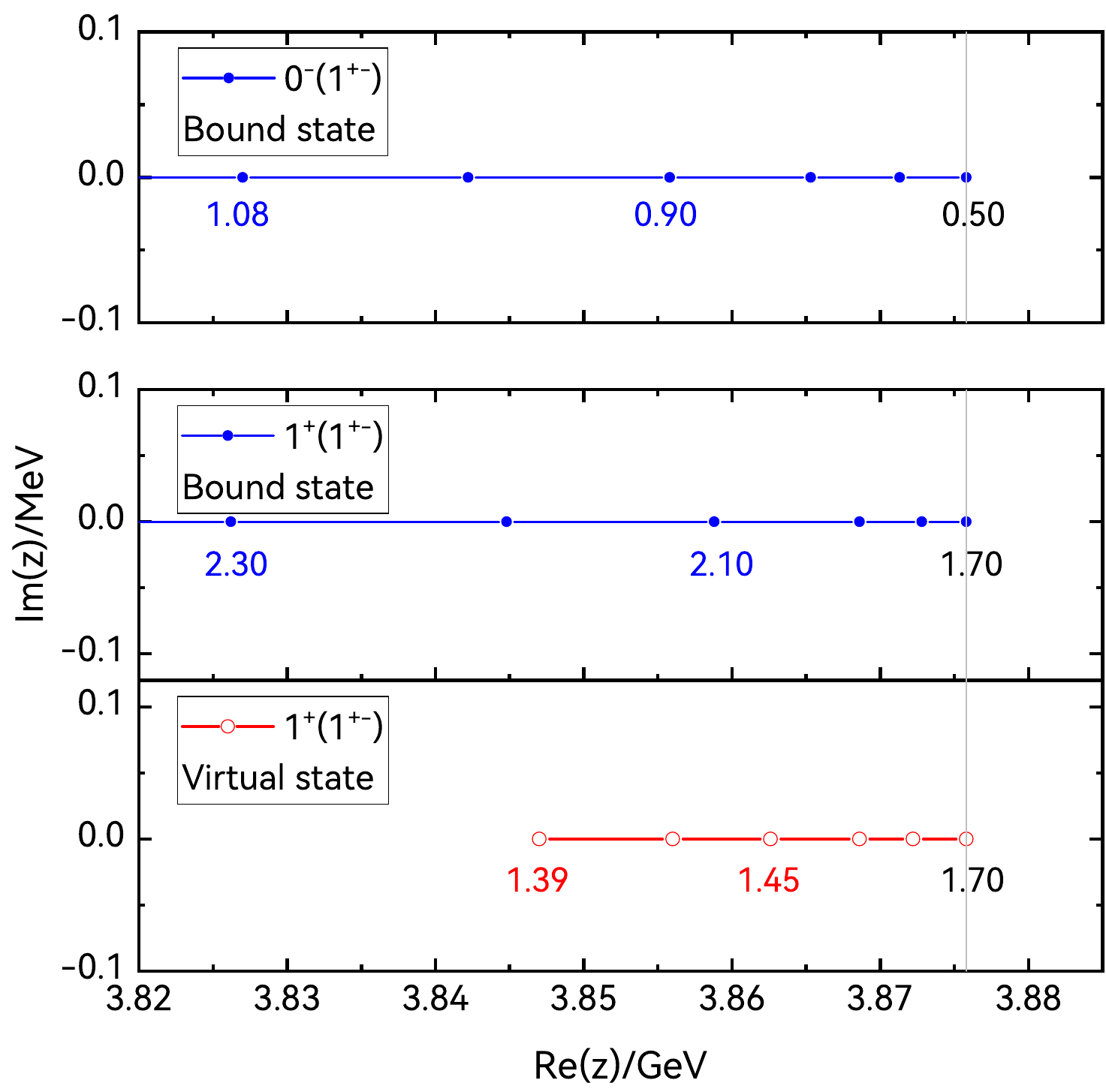}
\caption{Pole trajectories for the isoscalar bound (upper panel), isovector bound (middle panel) and isovector virtual (lower panel) states with quantum numbers $1^{+-}$. For the isoscalar state $0^{-}(1^{+-})$, the corresponding cutoff parameters from left to right are 1.08, 1.00, 0.90, 0.80, 0.70, and 0.50 GeV. For the isovector state $1^{+}(1^{+-})$, the bound state pole positions correspond to cutoff values (from left to right) of 2.3, 2.2, 2.1, 1.9, 1.8, and 1.7 GeV. The virtual state pole positions correspond to cutoff values (from right to left) of 1.70, 1.56, 1.50, 1.45, 1.41, and 1.39 GeV. The gray solid line indicates the $D\bar{D}^*$ threshold.}
\label{1+-}
\end{figure}

In Fig.~\ref{1+-}, the pole trajectories for the $J^{PC} = 1^{+-}$ channel are
shown. The trajectory for the isoscalar state is presented in the upper panel,
where only bound states are observed. The pole first appears at the threshold
when the cutoff $\Lambda$ is around 0.5~GeV. As the cutoff increases, the
attraction becomes stronger, causing the pole to move along the real axis on the
first Riemann sheet, reaching approximately 50~MeV below the threshold
when the cutoff is reduced to around 1.08~GeV, and continues to shift further as
the cutoff decreases.

The middle and lower panels of Fig.~\ref{1+-} show the pole trajectories of the
isovector state with quantum numbers $1^{+-}$. When the cutoff is very
large, corresponding to a strong attractive interaction, the pole appears deep in
the first Riemann sheet, far below the $D\bar{D}^*$ threshold. As the cutoff
decreases, the pole moves upward along the real axis toward the threshold. As
shown in the middle panel, the pole reaches a position approximately 50MeV below
the threshold when the cutoff is reduced to 2.30~GeV. Continuing to lower the
cutoff, the pole gradually approaches the threshold and reaches it at around
1.70~GeV.  Further reduction of the cutoff below 1.70~GeV weakens the interaction,
causing the pole to transition into the second Riemann sheet, where it becomes a
virtual state. This virtual state then moves away from the threshold as the
cutoff continues to decrease, eventually disappearing at a pole position of
approximately 3.847~GeV when the cutoff is reduced to 1.39~GeV.  Thus, depending
on the value of the cutoff, the isovector state with quantum numbers $1^{+-}$
can appear either as a bound state or as a virtual state. The presence of a
virtual state provides a plausible interpretation for the $Z_c(3900)$, which
manifests as a peak above the $D\bar{D}^*$ threshold, as discussed in
Refs.~\cite{He:2017lhy,Nakamura:2023obk}.

\subsection{P-wave pole trajectory}

\begin{figure}[h!]
\includegraphics[bb=0 0 1000 510,clip,scale=0.37]{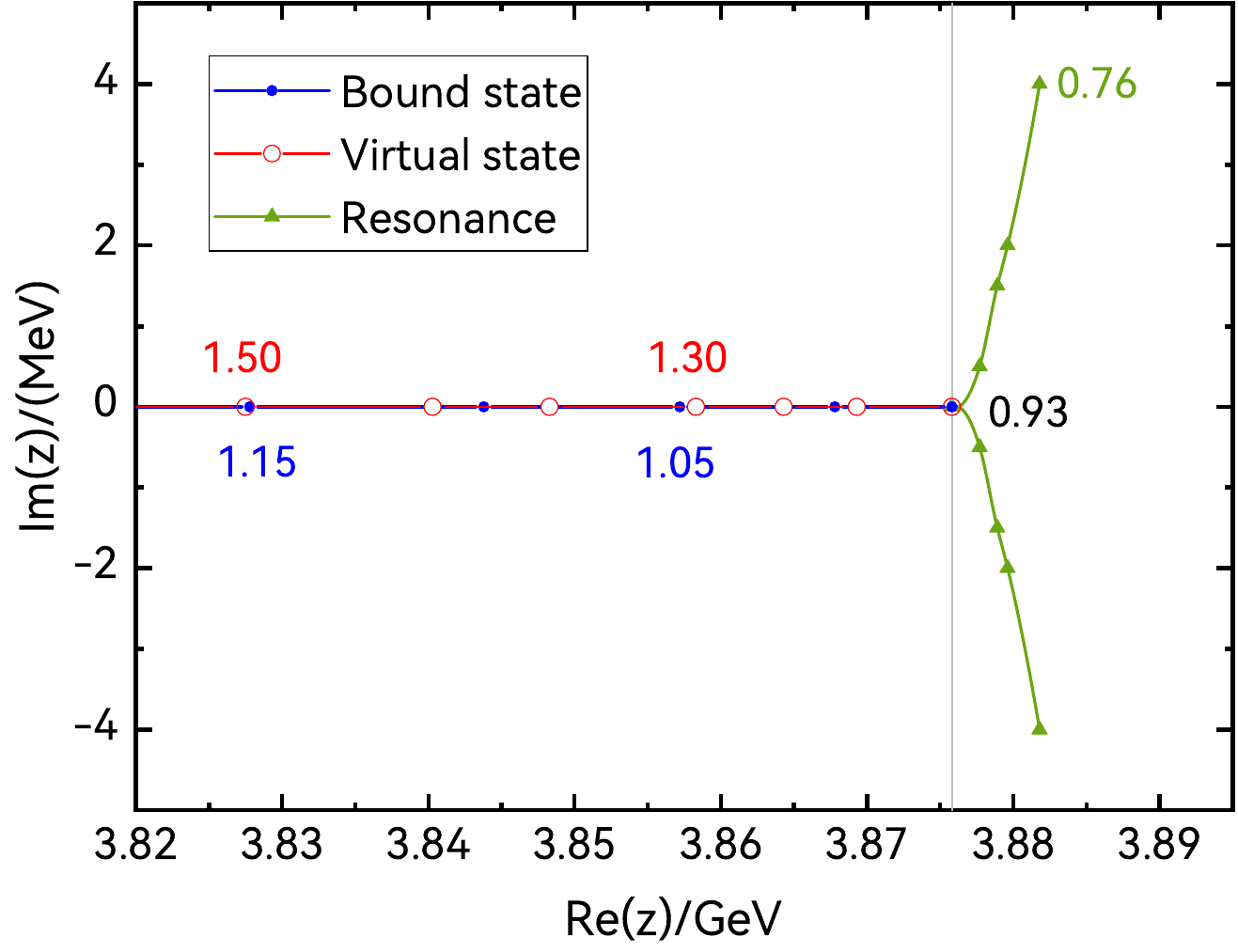}
\caption{Pole trajectory of the state with quantum numbers $0^{+}(0^{-+})$. The blue line represents the bound state on the first Riemann sheet, with the corresponding cutoff parameters (from left to right) being 1.15, 1.10, 1.05, 1.00, and 0.93~GeV. The red line denotes the virtual state on the second Riemann sheet, with cutoff values (from left to right) of 1.50, 1.45, 1.40, 1.30, 1.20, 1.10, and 0.93~GeV. The green line corresponds to the pair of conjugate poles in the complex plane, with cutoffs (from left to right) of 0.93, 0.89, 0.85, 0.81, and 0.76~GeV. The gray solid line indicates the $D\bar{D}^*$ threshold.}
\label{0-+}
\end{figure}

In this subsection, we present the results for the $P$-wave $D\bar{D}^*$
interaction, where only two isoscalar states are found, with quantum numbers
$0^{-+}$ and $1^{--}$. Figure~\ref{0-+} displays the evolution of the pole
positions for the $0^{+}(0^{-+})$ state. The results show that in the
single-channel $D\bar{D}^*$ interaction, both bound and virtual states can appear
on the first and second Riemann sheets, respectively, when the cutoff parameter
is below 0.93~GeV. These poles originate far from the threshold and gradually
move toward it as the interaction becomes weaker.  As shown in the figure, the
bound and virtual states approach the threshold along the real axis and enter
the range of interest, within 50~MeV below the threshold, at cutoff values of
1.15~GeV and 1.50~GeV, respectively. Although the two poles move at different
rates, they both reach the threshold simultaneously at a cutoff of 0.93~GeV,
which corresponds to the branch point connecting the two Riemann sheets.  If the
cutoff is further reduced, the poles cross the threshold and enter the complex
energy plane on the second Riemann sheet. In this region, they form a pair of
complex conjugate poles, departing from the real axis, and eventually disappear
at a cutoff of approximately 0.76~GeV, with a complex energy of $3.882 \pm
0.004$~GeV. This trajectory is consistent with the behavior discussed in
Ref.~\cite{Hanhart:2014ssa}.

\begin{figure}[h!]
\includegraphics[bb=0 0 1000 510,clip,scale=0.37]{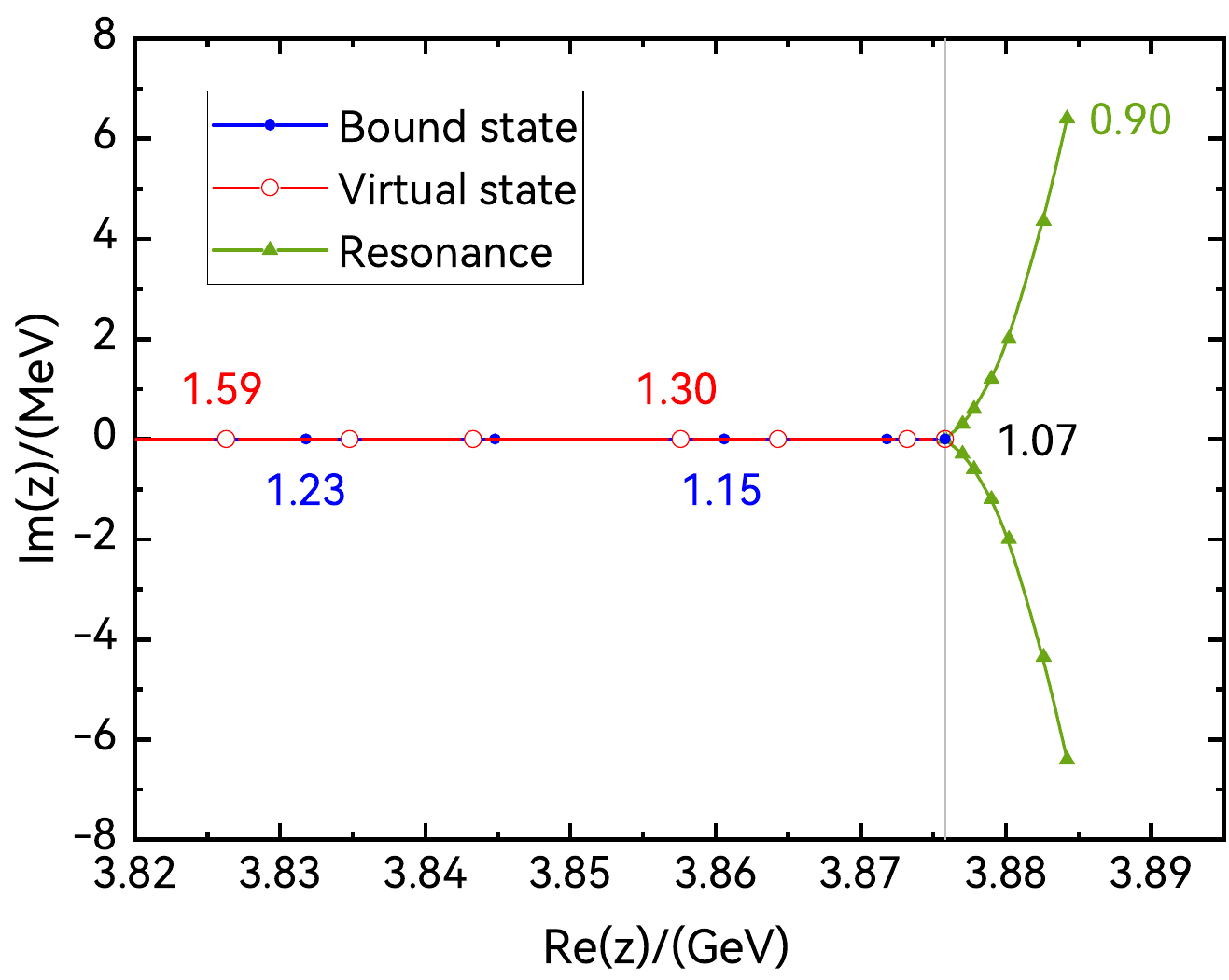}
\caption{Pole trajectory of the state with quantum numbers $0^{-}(1^{--})$. The blue line represents the bound state, with the corresponding cutoff parameters (from left to right) being 1.23, 1.20, 1.15, 1.10, and 1.07 GeV. The red line represents the virtual state, with cutoff parameters of 1.59, 1.50, 1.40, 1.30, 1.20, 1.10, and 1.07 GeV (from left to right). The green line represents the resonance trajectory, with cutoff parameters of 1.07, 1.05, 1.03, 1.01, 1.00, 0.95, and 0.90 GeV (from left to right). The gray solid line indicates the $D\bar{D}^*$ threshold.}
\label{0(1--)}
\end{figure}

Figure~\ref{0(1--)} shows the evolution of the pole positions for the state with
quantum numbers $I^{G}(J^{PC})=0^{-}(1^{--})$ arising from the $D\bar{D}^{*}$
interaction. Similar to the $0^{+}(0^{-+})$ case, it is evident that by
appropriately tuning the cutoff parameter $\Lambda$, both a bound state and a
virtual state can be generated below the threshold on the first and second
Riemann sheets, respectively. As the cutoff decreases, the two poles move along
the real axis on their respective Riemann sheets at different rates but reach
the branch point simultaneously at a cutoff of 1.07~GeV.  Upon further reduction
of the cutoff, weakening the interaction, the poles enter the complex energy
plane in the second Riemann sheet as a pair of complex conjugate poles, forming
a resonance. These poles eventually disappear at a cutoff of approximately
0.90~GeV, with a complex energy of $3.884 \pm 0.006$~GeV. The newly observed
structure $G(3900)$, located near the $D\bar{D}^*$ threshold and carrying the
quantum numbers $I^{G}(J^{PC})=0^{-}(1^{--})$, can thus be interpreted as a
molecular state dynamically generated from the $P$-wave $D\bar{D}^{}$
interaction.

\section{Summary} \label{Sec: Summary}

\begin{figure}[h!]
\includegraphics[bb=0 0 1000 490,clip,scale=0.386]{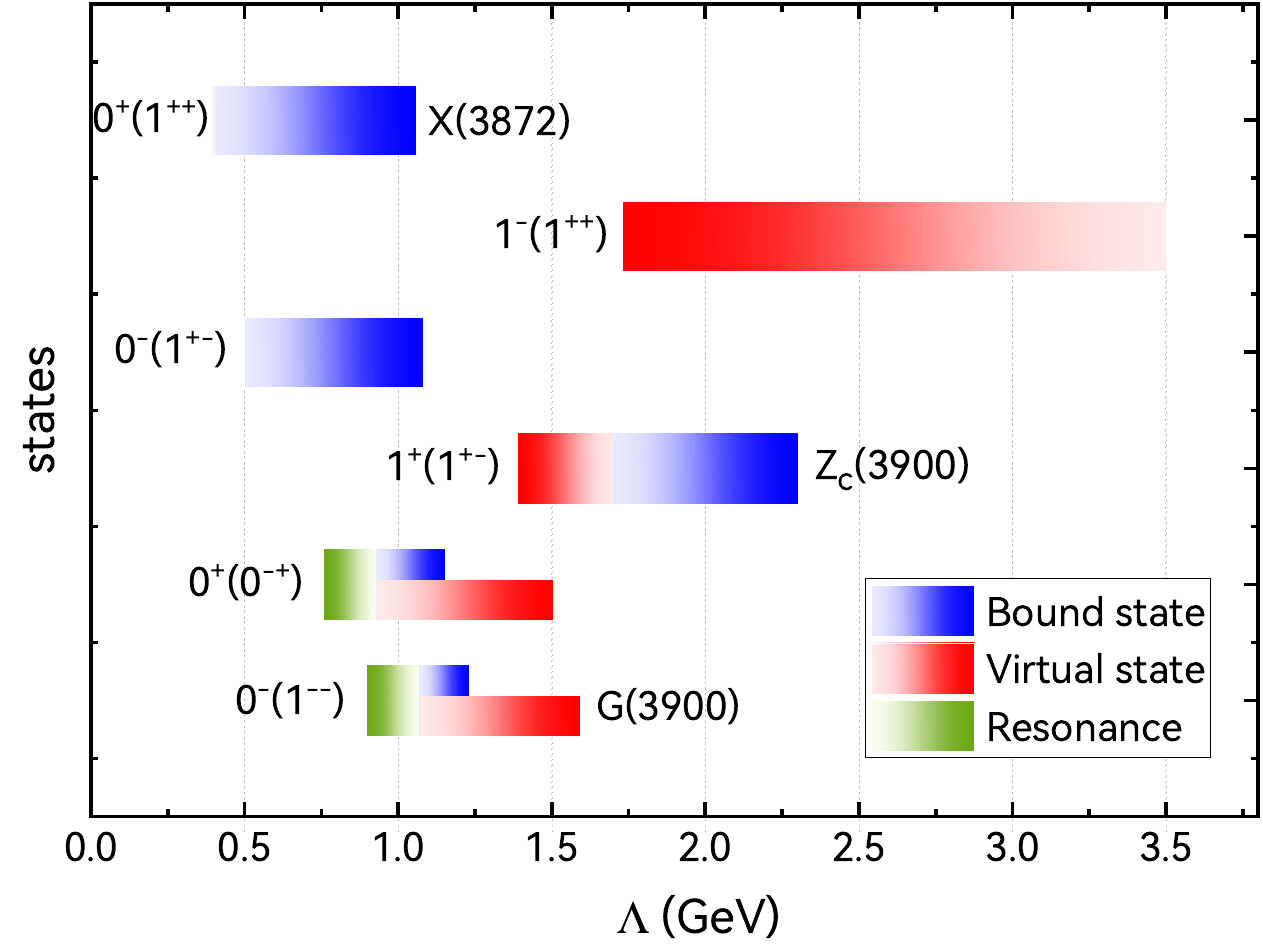}
\caption{Summary of the pole trajectories as a function of the cutoff parameters. Lighter colors represent poles located closer to the $D\bar{D}^*$ threshold.}
\label{sum}
\end{figure}

In the present work, we conduct a systematic investigation of the molecular
states arising from the $D\bar{D}^*$ interaction by analyzing their pole
trajectories. Within the framework of the qBSE, the interaction potentials are constructed based on heavy quark and
chiral symmetries. The poles of the scattering amplitude are examined to
identify possible bound states, virtual states, and resonances, as summarized in
Figure~\ref{sum}. 

For the states generated from the $S$-wave interaction, the isoscalar states
with quantum numbers $1^{++}$ and $1^{+-}$ can only appear as bound states by
adjusting the strength of the attraction through variations in the cutoff. The
$1^{++}$ state, in particular, can be associated with the well-known $X(3872)$.
Moreover, an isovector partner of the $X(3872)$ is also predicted to exist,
but only as a virtual state. This result aligns with the analysis in
Ref.~\cite{Ji:2025hjw}.

Since virtual states can exist independently without being accompanied by bound states as $1^-(1^{++})$ state, the existence of virtual states is often overlooked, as bound states tend to
attract more attention. However, our findings suggest that bound states, virtual
states, and resonances should all be considered when studying hadronic
molecules. Notably, virtual states, which can also account for observed structures
above thresholds, as discussed in Refs.~\cite{Du:2022jjv,He:2017lhy}, can produced solely. In
particular, our results support the interpretation of the isovector $1^{+-}$
state, which may appear as either a bound or virtual state, as a possible
explanation for the $Z_c(3900)$, observed above the $D\bar{D}^*$ threshold.

The spin parity of a $D\bar{D}^*$ system with negative parity must correspond to
a $P$-wave state. If the recently observed isoscalar $G(3900)$ state with
quantum numbers $1^{--}$ is interpreted as a molecular state arising from the
$D\bar{D}^*$ interaction, it should indeed be assigned as a $P$-wave state. Our
results support the existence of two isoscalar $P$-wave states with quantum
numbers $0^{-+}$ and $1^{--}$, with the latter being associated with the
$G(3900)$. For both states, the variation of the cutoff results in different
possible configurations: bound states, virtual states, or resonances.

In the current work, we also study the behavior of pole trajectories. For the
$S$-wave, only bound states and virtual states are produced for the $D\bar{D}^*$ interaction, with no resonances
observed. In contrast, for the $P$-wave, all three types of poles, bound states,
virtual states, and resonances, are produced as the cutoff varies. The
trajectories for the two states are in good agreement with the theoretical
discussions presented in Refs.~\cite{Hanhart:2014ssa, Taylor:1972pty}. Thus, the
current results provide a concrete example of the pole trajectory for $P$-wave
interactions.

As shown in Fig.~\ref{sum}, the cutoff ranges for which solutions exist in
different cases do not always overlap. In the present work, we focus on
single-channel calculations to avoid the complexities associated with
coupled-channel dynamics and to provide a clearer understanding of the pole
trajectories. In principle, the inclusion of coupled-channel effects may shift
the pole positions through inter-channel interactions. However, such effects do
not necessarily resolve the observed mismatch in cutoff values across different
channels.  For example, if a single cutoff reproduces the $X(3872)$ but fails to
generate the $Z_c(3900)$ or $G(3900)$, this would suggest that these states
cannot  all be interpreted as molecular states within the same parameter setup in
a coupled-channel framework.   A definitive conclusion
requires a full coupled-channel analysis

In addition, we emphasize that the current OBE model does
not include explicit local four-point contact interactions, which are essential
ingredients in effective field theory  descriptions of short-range
dynamics. Although heavy meson exchanges such as $\rho$, $\omega$, and $J/\psi$
introduce short-range interactions that partially mimic contact-like contributions,
the absence of genuine contact terms restricts the interaction to a particular
slice of the full parameter space. As a result, the cutoff dependence in our
analysis provides only a lower bound on the true systematic uncertainty, and a
more complete description would require incorporating contact terms and
performing a consistent renormalization procedure.

We present a systematic analysis of the $S$- and $P$-wave $D\bar{D}^*$
interactions using a single-channel OBE framework,
extracting the corresponding pole trajectories and identifying the conditions
under which molecular states such as the $X(3872)$, $Z_c(3900)$, and $G(3900)$
may emerge. While the present model does not include explicit short-range
contact interactions and is restricted to single-channel dynamics, it
effectively captures the essential features of long-range forces and offers
valuable insight into the formation mechanisms of near-threshold states. These
results provide a solid baseline for future investigations incorporating
coupled-channel effects and a more complete treatment of short-range dynamics,
with the goal of achieving a more definitive understanding of the $S$- and
$P$-wave $D\bar{D}^*$ interactions and their implications for hadronic molecular
states.

\vskip 10pt \noindent {\bf Acknowledgement} This project is supported by the National Science
Foundation of China (Grant No. 12475080)

\end{document}